# Modeling of the equilibrium component of the stress tensor of filled elastomeric materials with taking into account the Mullins softening effect

*Mokhireva K. A.[1], Svistkov A. L.[1,2]*

[1] *Institute of Continuous Media Mechanics of the Ural branch of the Russian Academy of Science, Akademika Koroleva str. 1, Perm, Russia, 614013*

[2] *Perm State University, Bukireva str. 15, Perm, Russia, 614990*

**Abstract**

Elastomers are viscoelastic materials and their properties significantly depend on the loading rate. The actual stress experienced by these materials is the sum of equilibrium and dissipative (inelastic) terms. At very low loading rates we can eliminate the significant influence of time effects and model the material as hyperelastic. In this paper, the features of the experimental determination and subsequent mathematical description of equilibrium stresses are considered. Verification of the proposed equations has been carried out for a series of experiments - cyclic uniaxial tests of samples of materials on the basis of the same matrix, but with different filler contents and under different maximum degrees of deformation.

**Introduction**

Rubber products are widely used in different fields of modern world due to their physical and mechanical properties. However, the complex nature of their behavior (finite deformations, dependence of polymer properties on strain rate, hysteresis losses, stress relaxation, softening effect, etc.) causes difficulties with creation of a system of constitutive equations for this class of materials. In order to take into account the effect of loading rate on the elastomers behavior, two approaches are usually used: integral and differential [1, 2]. Despite the differences in the methods of constructing equations, both approaches have one similar feature. In each approach we can directly determine equilibrium (elastic) stresses, which are independent of time processes.

As a rule, most researchers do not pay special attention to the description of the elastic component and define it with the help of "simple" elastic potentials. Such simplification is not correct when modeling the properties of high-filled rubbers under conditions of large deformations (more than 100%). In addition, it has been long established that the material softens (stresses fall) under reloading - the Mullins effect [3]. This feature should also be taken into account when describing the experimental data. In this article, the peculiarities of the experimental procedure aimed to determine the equilibrium component of stresses are considered and equations describing the elastic component of the stress tensor are proposed.

**Elastic properties. Formulation of the elastic potential**

As it has been mentioned before, the viscoelastic behavior of the elastic material is described by the models of integral and differential forms [1, 2], of which the differential models are most frequently used. This is due to the simplicity and clarity of the tensor functions used in the equations, which reflect the physics of the processes occurring during deformation. It is proposed to consider the free energy in the form of a sum of equilibrium and nonequilibrium (dissipative) parts [4]:

$$\mathbf{T} = \mathbf{T}_e + \mathbf{T}_d, \qquad (1)$$

where $\mathbf{T}_e$ is the actual equilibrium stress at given deformation after the completion of all transient processes, and $\mathbf{T}_d$ is the dissipative stress (difference between real $\mathbf{T}$ and equilibrium $\mathbf{T}_e$ stresses).

Let us consider the construction of the constitutive equations for the equilibrium component $\mathbf{T}_e$ only. Usually, this component of the stress tensor is determined by specifying an elastic potential, the form of which can be chosen from the enormous variety of potentials already proposed [3, 5]. The presence of a large number of representations for describing the elastic deformation energy can be attributed to attempts to specify an optimal form providing a reliable description of the experimental data regardless of the material and the way of its loading. Furthermore, this potential should include a small number of constants to be determined.

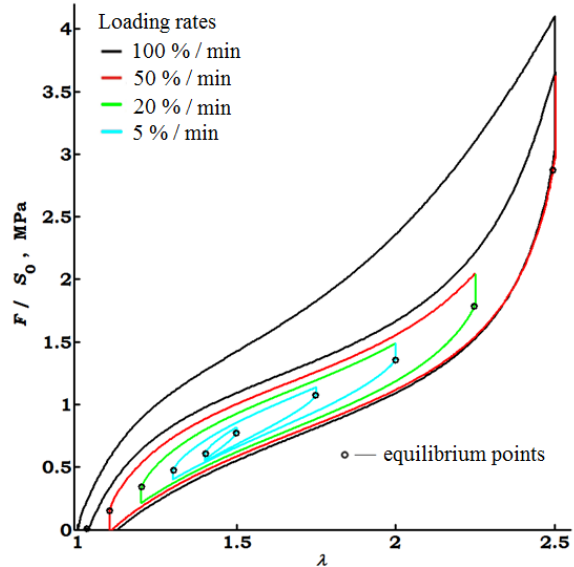

Figure 1. Uniaxial loading of the sample made of the material filled with 50 phr carbon black N220. From one sample, it is possible to obtain the data on the viscoelastic behavior of the material, residual strains, the Mullins softening effect, and equilibrium points at different degrees of the sample elongation. *F* - acting force, *S*0 - initial cross section of the sample, $\lambda$ - sample stretch ratio.

Before choosing the shape of the potential for describing the elastic properties of elastomers, it is necessary to determine the method of finding elastic (equilibrium) stresses. To this end, we have performed the experiments, in which the samples were subjected to uniaxial cyclic loading according to the loading scheme described in patent [6] and illustrated in Figure 1. After each loading and unloading of the sample, a 20-minute delay was set for the first deformation cycle and a 10-minute delay for the subsequent cycles. After 20- or 10-minute "rest" of the material, the curve breakpoints (equilibrium points) were formed at each of the unloading and loading sections (Figure 1). In our opinion, these points characterize adequately the state of the material under equilibrium (quasistatic) tension.

This type of the test was performed for the samples based on the same matrix with different filler concentration. As an elastomeric base, we have used styrene-butadiene rubber SBR-1502, which has found wide application in the tire, rubber, cable and footwear industries. The filler was carbon black ISAF N220 with spherical particles of low density $\rho = 1.8 \, \text{g} / \text{cm}^3$. The filler content was as follows: 1) 10 parts by weight of carbon black per 100 parts by weight of rubber, i.e. 10 phr, 2) 30 phr, 3) 50 phr. Accordingly, the volume fraction occupied by the filler in the composite for each case was: 1) $\varphi \approx 4.8\%$, 2) $\varphi \approx 13.3\%$ and 3) $\varphi \approx 20.4\%$.

Since the material behavior depends on the degree of loading, in order to determine the universal shape of the potential and to more accurately specify the values of the constants of this potential, we performed a series of tests on cyclic loading of samples at different maximum stretch ratios: a) $\lambda = \dfrac{l}{l_0} = 2$, b) $\lambda = 2.5$ c) $\lambda = 3.5$, d) $\lambda = 4$.

By analyzing the obtained experimental data, we have defined the potential describing the elastic strain energy density as follows:

$$\omega = a\,C \ln(I_1) - C \ln\left(1 - \frac{I_1}{I_*}\right) + \text{const}, \qquad (2)$$

where $a$, $C$ is the material constants, $I_1 = \lambda_1^2 + \lambda_2^2 + \lambda_3^2$, $\lambda = \dfrac{l}{l_0}$ is the stretch ratio, and $I_*$ is the limiting value for $I_1$. The material is assumed to be incompressible $(\lambda_1 \lambda_2 \lambda_3 = 1)$. The proposed potential is a combination of two parts: the first term for the initial loading segment and the second term for the sudden increase in stresses with increasing strains. The second component is the variation of the Ghent potential [5], where $I_*$ characterizes the limiting extensibility of the rubber network so that the value of this parameter depends on the degree of the sample elongation.

The use of this potential in calculating the equilibrium component of the stress tensor requires determination of two material constants and one parameter characterizing the behavior of elastomers. For this purpose, we have minimized the sum of squares of normalized deviations of the model data from the experimental results. Normalization of the differences between the numerical and experimental results is given via dividing by the value of the experimental data. Being subjected to loading, the material accumulates residual strains; the higher is the loading force, the larger are the residual strains. In order that the error of the model does not exceed 5%, especially in the cases of highly filled systems (filler content is more than 30 phr) under the strains exceeding 250%, it is necessary to shift the curve along the X-axis by the value equal to the residual strain. Since the model has been developed under the assumption of a multiplicative decomposition of the deformation gradient into elastic and inelastic parts, the following condition is valid: $\lambda = \lambda_e \lambda_p$, where $\lambda$ is the total stretch ratio, $\lambda_e$ is the elastic (equilibrium) stretch ratio and $\lambda_p$ is the inelastic (plastic) stretch ratio. Therefore, the displacement of the curve along the X-axis with the aim to describe the equilibrium curve of the stress-softened material behavior is expressed as:

$$\lambda_e = \frac{\lambda}{\lambda_p}. \tag{3}$$

In further equations and calculations, instead of $\lambda$, we use $\lambda_e$, and $\lambda_i^e$ $(i=1,2,3)$ correspond to the elastic components of the stretch ratio found from (3).

For constructing theoretical "stress-stretch ratio" curves, elastic stresses are calculated based on the formula:

$$\sigma_1^e = \frac{F_e}{S_0} = \lambda_1^e \frac{\partial \omega_e}{\partial \lambda_1^e} - p = 2\lambda_1^e C \left(\lambda_1^e - \frac{1}{(\lambda_1^e)^2}\right)\left(\frac{\alpha}{I_1^e} + \frac{I_*}{(I_* - I_1^e)}\right), \tag{4}$$

where $p$ is the Lagrange parameter defined by the condition $(\sigma_2^e = \sigma_3^e = 0)$. Thus, according to equation (3), the constants $\alpha$, $C$ and the parameter $I_*$ determine the stretching of the curve along the Y-axis, and only $I_*$ specifies the stretching along the X-axis. We assume that the constant $\alpha$ is invariable for all materials to exclude the mutual influence of $\alpha$ and $C$ on each other. We also assume that $\alpha = 5$, then, minimizing the sum of squares of the deviations between the numerical and experimental results, we arrive at the conclusion that the constant $C$ varies insignificantly within the framework of one material. Let this constant be the same in the calculations and have a certain value for each material.

Changes in the model constants $(\alpha, C)$ and the parameter $I_*$ for the examined elastomer samples is given in Table 1.

Table 1. The constants and parameters of the developed model for different materials

| Material | Total stretch ratio $\lambda$ | Constant $\alpha$ | Constant $C$ | Parameter $I_*$ |
|---|---|---|---|---|
| filler volume fraction $\varphi \approx 4.9\%$ | 2 | 5 | 0.093 | 16.857 |
| | 2.5 | | | 18.71 |
| | 3.5 | | | 23.805 |
| | 4 | | | - |
| filler volume fraction $\varphi \approx 13.3\%$ | 2 | | 0.106 | 9.534 |
| | 2.5 | | | 11.134 |
| | 3.5 | | | 15.684 |
| | 4 | | | 19.557 |
| filler volume fraction $\varphi \approx 20.4\%$ | 2 | | 0.125 | 7.329 |
| | 2.5 | | | 8.857 |
| | 3.5 | | | 12.828 |
| | 4 | | | 16.406 |

Figure 2 presents examples showing how the experimental results can be approximated by the proposed model in accordance to the values from Table 1. It is seen that the theoretical curves approximate the experimental data well.

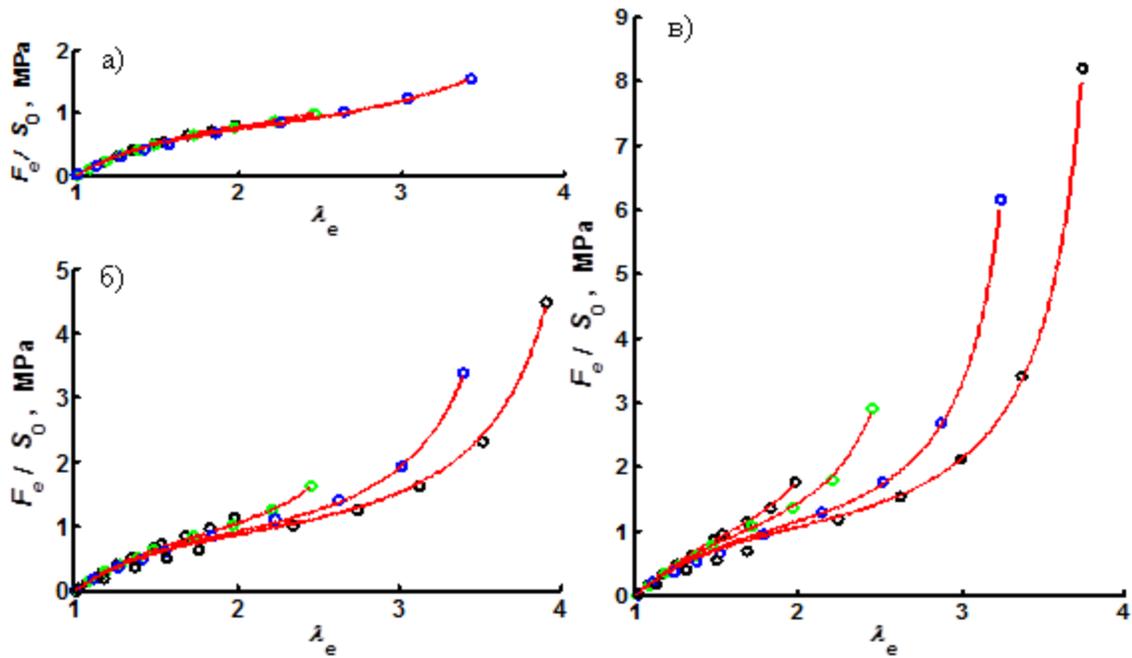

Figure 2. Plots of the experimental data (points) and the theoretical curves (lines) for materials with filler content: a) 10 phr; b) 30 phr; c) 50 phr

Based on the obtained constitutive equations (8.2) and the calculated results (Table 8.1), we put forward some assumptions. Firstly, for all materials the constant $\alpha$ is regarded as an invariable value and, therefore, characterizes the properties of the material matrix and, from the physical point of view, can be considered as the elongation ratio of polymer chains. The constant $C$ increases with increasing filler concentration in the elastomer and, in fact, shows the degree of reinforcement depending on the degree of filler concentration. Secondly, the parameter $I_*$, as already mentioned, reflects the limiting value of the network extensibility. In the case of the material with a filler content of 10 phr, the value of this constant has exceeded the maximum permitted value and because of this the results for the sample stretched to 300% are absent; a break occurs.

Using the results obtained in our investigation, we have derived an elastic potential (2) that involves constants, which are able to characterize the elastomer properties. With this potential, we can provide a relative evaluation of the reinforcement degree of elastomers (the smaller is the constant value, the less is the active filler concentration in the material). The parameter $I_*$ allows one to determine the elongation degree of the sample (the higher is its value, the higher is the stretch ratio and the greater is the probability of a break).

**Analysis of the relations between the obtained constants. The Mullins softening effect.**

Let us consider the dependence of the constant $C$ and the parameter $I_*$ of the elastic potential on the choice of the material (the degree of its filling) and the peculiarities of the testing procedure (different maximum strain levels).

The calculations of the model constant $C$ (4) have shown that its value increases with higher concentration of active filler particles in the material. Comparison of the obtained calculations with the experimental data indicates that the relationship between the constant $C$ and the filler concentration $\varphi$ is linear (Figure 3) and can be determined as

$$C = C_1 + C_2\varphi = 0.082 + 0.2\varphi, \tag{5}$$

where $C_1$ and $C_2$ are the constants, and $\varphi$ is the volume fraction of filler in the material.

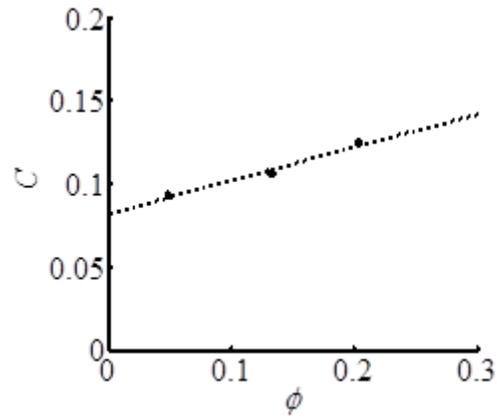

Figure 3. Linear relationship between the material constant $C$ and the filler volume fraction $\varphi$ of the composite

Besides, it has been found that the parameter $I_*$ depends on the stretch ratio, and the parameter increases as the prescribed maximum strain grows (Figure 4). For each of the materials, this relationship is inherently linear:

$$I_* = C_3 + C_4^* \max\left(I_1^e\right) = \begin{cases} 1.9 + \max\left(I_1^e\right), & \varphi \approx 0.049 \\ 3.8 + \max\left(I_1^e\right), & \varphi \approx 0.133 \\ 11.4 + \max\left(I_1^e\right), & \varphi \approx 0.205 \end{cases}. \qquad (6)$$

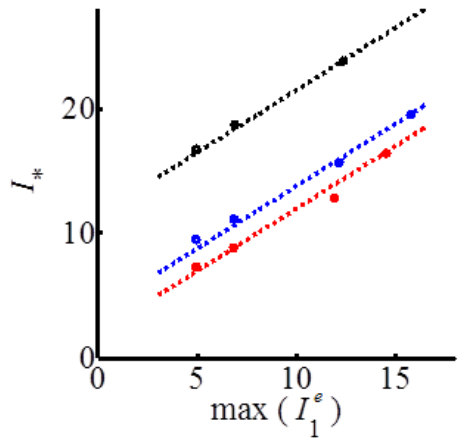

Figure 4. Linear relationship between the model parameter $I_*$ and the stretch ratio

It follows from the previous calculations of the parameter $I_*$ that the constant $C_4^* = 1,$ and therefore it is not further considered for the term $\max(I_1)$ in equations (6), but $C_3$ depends on the variable $\varphi$ (Figure 5):

$$C_3 = C_4 + C_5\varphi^2 = 1.5 + 200\varphi^2. \tag{7}$$

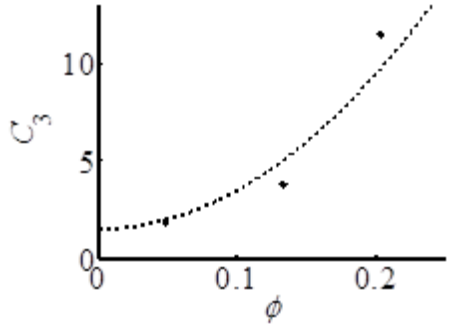

Figure 5. Power-law dependence of the constant $C_3$ on the filler volume fraction $\varphi$ of the material

Taking into account equations (5), (6) and (7), potential (2) can be written in the generalized form as

$$\omega_e = (C_1 + C_2\varphi)\left[\alpha \ln(I_1^e) + \ln\left(1 - \frac{I_1^e}{C_4 + C_5\varphi^2 + \max(I_1^e)}\right)\right] + \text{const}. \tag{8}$$

Thus, equation (8) allows us to find and describe the entire spectrum of the equilibrium curves of the behavior of stress-softened materials, regardless of their filler concentration in them and the maximum stretch ratio given in the experiment.

The equilibrium strain curves received from the experiments show that the material has softened after being subjected to loading, and this reduces the material stiffness during the repeated loading cycles (the Mullins effect) [7, 8]. From the practical point of view and modeling the softening of elastomers under deformations is an undesirable effect. If the material is able to soften completely, then there is no need to take this effect into account in calculations, but in practice the material is able to recover its properties with time partially or almost completely, especially during "rest" at high temperatures [9]. Furthermore, in real conditions rubber products experience nonuniform loads and, consequently, the material exhibits the nonuniform softening effect. Therefore, within the framework of numerical and finite-element calculations it is necessary to take into account the Mullins effect and to determine the behavior of the non-softened material.

In our case, the relationship represented by equation (8) with consideration of (4) permits us to determine the equilibrium stress-strain curves of the non-softened material. Figure 6 shows the equilibrium tensile curves of the material deformed for the first time (blue lines) and the set of equilibrium curves of the stress-softened material (red lines). These lines are the stain-stress curves of the material that experienced a total stretch of 100%, 150%, 250% and 300%, respectively. It is worthy to note that the current deviations between the equilibrium tensile curves of the stress-softened and non-softened materials with the test data indicate only possible experimental and computational errors.

As a result, we have derived the constitutive equation (8) for describing and finding the elastic properties of materials.

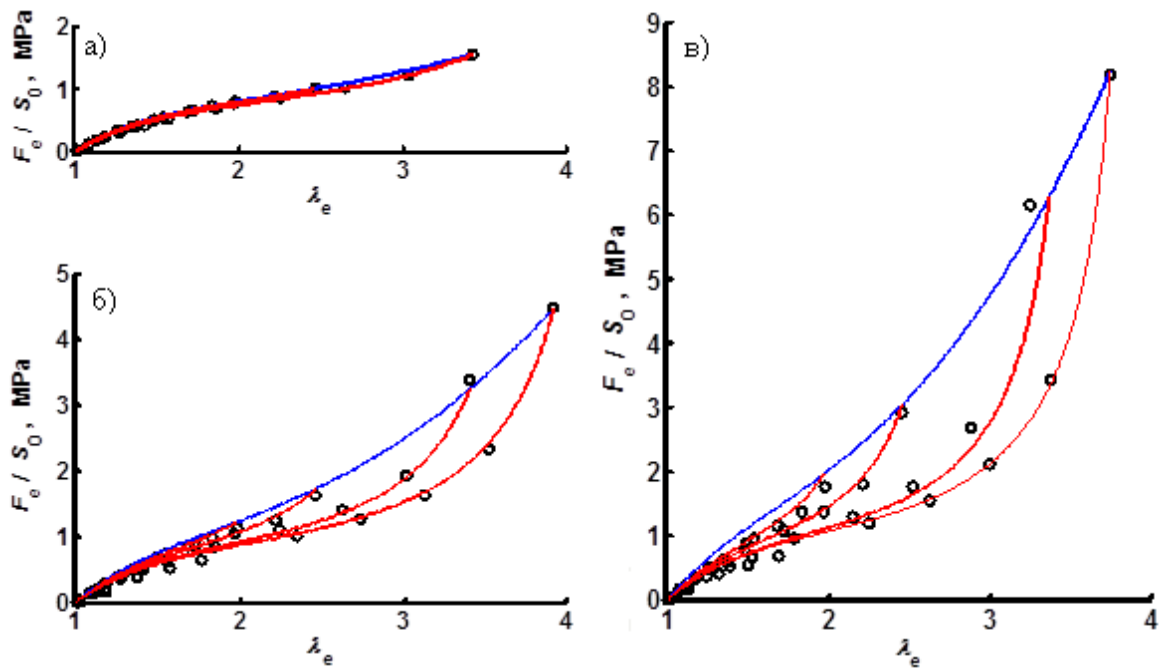

Figure 6. Curves of the elastic behavior of the material deformed for the first time (blue lines) and the stress-softened material (red lines) with filler concentrations: a) 10 phr; b) 30 phr; c) 50 phr

**Conclusions**

A possible way of determining and describing the elastic component of the stress tensor, which is needed for modeling the viscoelastic behavior of elastomers, is considered.

An experimental method to specify the equilibrium curves is presented.

We have developed a five-constant elastic potential that involves the constants characterizing the properties of elastomers depending on their structure (the amount of the filler introduced) and the peculiarities of the testing procedure (the degree of loading)

The efficiency of the proposed model has been validated by comparison with the experimental data for materials based on the same matrix with different filler contents.

**Acknowledgements**

The study was supported by RFBR (grant numbers 16-08-00914, 17-08-01118)